\documentstyle[twoside,fleqn,espcrc2,epsf]{article}

\newcommand{\AmS}{{\protect\the\textfont2
  A\kern-.1667em\lower.5ex\hbox{M}\kern-.125emS}}

\title{Quenched Light Hadron Spectrum and Decay Constants using Improved Wilson
Fermion Actions}

\author{UKQCD Collaboration, presented by Richard
Kenway\address{Department of Physics \& Astronomy, The University of
Edinburgh, Edinburgh EH9 3JZ, Scotland}\thanks{Work supported by EPSRC
grant GR/K41663 and PPARC grants GR/J98202, GR/K54601, GR/K55745.}}
       
\begin{document}

\begin{abstract} 
We compare results obtained using the Sheikholeslami-Wohlert (SW)
fermion action with tree-level and tadpole-improved coefficients for
$5.7\le\beta\le 6.2$. 
\end{abstract}

\maketitle

\section{LATTICE PARAMETERS}

The use of a tadpole-improved coefficient~\cite{lepage} for the clover
term in the SW fermion action,
\begin{eqnarray}
S_F 
&=& \sum_{xy}\bar{q}(x)
    \Big\{\big[1-\frac{ic\kappa}{2}\sigma_{\mu\nu}F_{\mu\nu}(x)\big]
    \delta_{xy}\nonumber\\
& & -\kappa\big[(1-\gamma_\mu)U_\mu(x)\delta_{x+\hat{\mu},y}\nonumber\\
& & +(1+\gamma_\mu)U_\mu^\dagger(y)\delta_{x-\hat{\mu},y}\big]\Big\}q(y),
\end{eqnarray}
where $c = 1/u_0^3$, $u_0 =
\langle\frac{1}{3}\mbox{Tr}\,\Box\rangle^{1/4}$, can be regarded as a
step towards full $O(a)$ improvement~\cite{alpha}.  We compare the
resulting light hadron quantities with those obtained using the
tree-level coefficient, $c=1$, for which discretisation errors are
$O(g^2a)$ in perturbation theory~\cite{heatlie}, and also with data
from GF11~\cite{gf11} corresponding to $c=0$, at a range of $\beta$
values, to seek indications of better scaling behaviour. 

Our data set comprises: at $\beta = 5.7$, Jacobi-smeared quark
propagators at two $\kappa$ values with $c=1/u_0^3$ and $c=1$ on 142
$16^3\times 32$ configurations; at $\beta = 6.0$, fuzzed
propagators~\cite{fuzzing} at three $\kappa$ values with $c=1/u_0^3$ on
499 $16^3\times 48$ configurations; and, at $\beta = 6.2$, fuzzed
(local) propagators at three (two) $\kappa$ values with $c=1/u_0^3$
($c=1$) on 130 (60) $24^3\times 48$ configurations.  We construct meson
correlators from all $\kappa$ combinations, but baryon correlators only
from degenerate combinations.  Consequently, we do not have enough
baryon data to perform a reliable chiral extrapolation.  Hadron masses
are obtained from multi-exponential fits to various combinations of
smeared and local correlators.  The results presented are from a
preliminary analysis of our current data (see also~\cite{lat95}); higher
statistics at $\beta=6.2$ will be available soon. 

\section{HADRON SPECTRUM}

\begin{figure}[htb]
\epsfxsize=0.4\textwidth
\epsffile{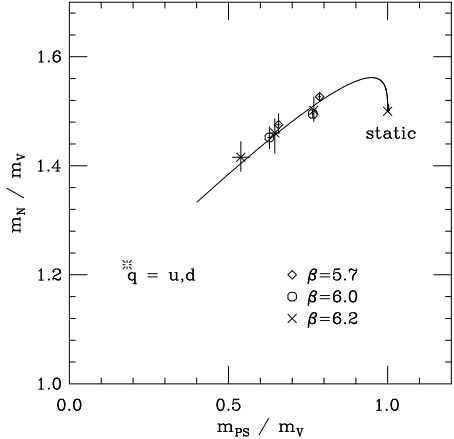}
\caption{Scaling at fixed \protect{$m_{PS}/m_V$} for \protect{$c=1/u_0^3$}.}
\label{fig:ed_scaling}
\end{figure}

The ratio of the nucleon to vector meson mass, $m_N/m_V$, at a fixed
ratio of the pseudoscalar to vector meson mass, $m_{PS}/m_V$, has a
noticeable dependence on $c$ at our largest lattice spacing.  The trend
is towards improved scaling for this ratio as $c$ is increased to the
tadpole-improved value.  Evidence for the latter is given in the
Edinburgh plot in Figure~\ref{fig:ed_scaling}. 

\begin{figure}[htb]
\epsfxsize=0.4\textwidth
\epsffile{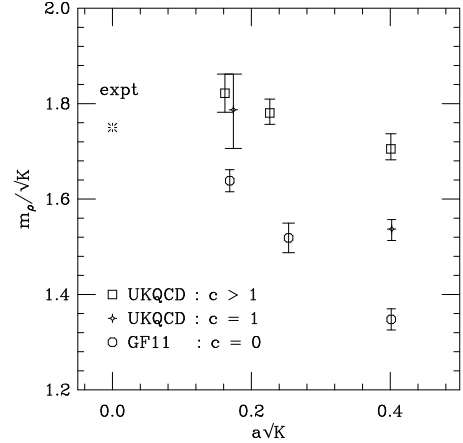}
\caption{\protect{$m_\rho$} in units of the string tension.}
\label{fig:m_rho}
\end{figure}

Linear extrapolation of data for the pseudoscalar meson mass to
$m_{PS}(\kappa_c,\kappa_c)=0$ for the tadpole-improved action gives
$u_0\kappa_c = 0.12347(3)$, 0.12224(1), and 0.12208(2), for $\beta =
5.7$, 6.0 and 6.2 respectively, in reasonable agreement with the
tree-level value of 0.125.  We define $\kappa_{ud}$, corresponding to
degenerate $u$ and $d$ quarks, at $m_{PS}/m_V = m_\pi/m_\rho$.  Then, as
shown in Figure~\ref{fig:m_rho}, our estimates for $m_\rho$ in units of
the square root of the string tension, $\sqrt{K}$, have decreasing
dependence on the lattice spacing, $a$, as $c$ is increased from 0 to
$1/u_0^3$, although the dependence is not removed entirely. 

\begin{figure}[htb]
\epsfxsize=0.4\textwidth
\epsffile{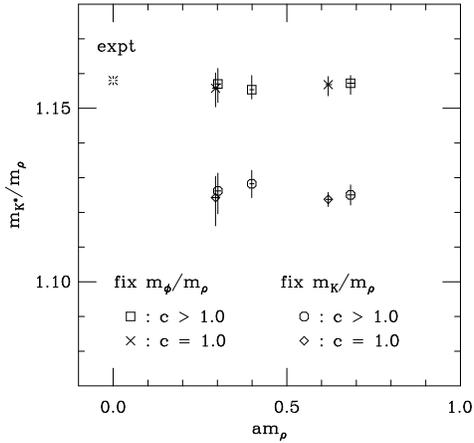}
\caption{\protect{$m_{K^\ast}/m_\rho$} for two definitions of $\kappa_s$.}
\label{fig:kaons}
\end{figure}

The $\kappa$ value corresponding to the strange quark mass, $\kappa_s$,
may be fixed from any one of the ratios $m_K/m_\rho$, $m_\phi/m_\rho$
and $m_{K^\ast}/m_\rho$.  In Figure~\ref{fig:kaons} we show the results
of fixing one of the first two ratios and calculating the third.
Evidently, the $K^\ast$ mass scales with the $\rho$ mass in both cases,
independently of whether $c=1$ or $1/u_0^3$, but different values are
obtained for $m_{K^\ast}/m_\rho$.  This is an indication that the
strange quark mass cannot be determined consistently in the quenched
approximation~\cite{rajan}.  We also find that
$J$~\cite{michael_lacock} is not changed by tadpole improvement, and
scales at a value inconsistent with experiment.

Finally, we observe that, at all three $\beta$ values, the magnitude of
the meson spin splitting, $m_V^2-m_{PS}^2$ in units of $m_K$ or
$m_\rho$, is insensitive to whether $c=1$ or $1/u_0^3$.

\section{DECAY CONSTANTS}

Our quark propagators are not `rotated', so $O(a)$-improved matrix
elements are constructed from improved quark fields:
\begin{eqnarray}
q^I
&=& \sqrt{2\kappa u_0}\Big[1 
    - \frac{\vec{D\hspace{-0.7em}/}}{2u_0}\Big]q\nonumber\\
&=& \sqrt{2\kappa u_0}\Big[1 
    + \frac{1}{2u_0}\Big(\frac{1}{2\kappa}-\frac{1}{2\kappa_c}\Big)\Big]q
\end{eqnarray}
where we have used the equations of motion and employed the
tadpole-improvement prescription~\cite{lepage}.  Thus, we take the
tadpole-improved pion decay matrix element to be~\cite{rowland}
\begin{eqnarray}
\langle 0|\bar{q}^I\gamma_4\gamma_5q^I|\pi\rangle\hspace{-1.5mm}
&=& \hspace{-1.5mm}Z_A\,2\kappa u_0\Big[1 
    + \frac{1}{u_0}\Big(\frac{1}{2\kappa}
    -\frac{1}{2\kappa_c}\Big)\Big]\nonumber\\
& & \hspace{-1.5mm}\times \langle 0|\bar{q}\gamma_4\gamma_5q|\pi\rangle
\end{eqnarray}
where we obtain $\kappa_c$ from $m_{PS}(\kappa_c,\kappa_c)=0$ and the
current normalisation, $Z_A$, from tadpole-improved one-loop
perturbation theory. 

\begin{figure}[htb]
\epsfxsize=0.4\textwidth
\epsffile{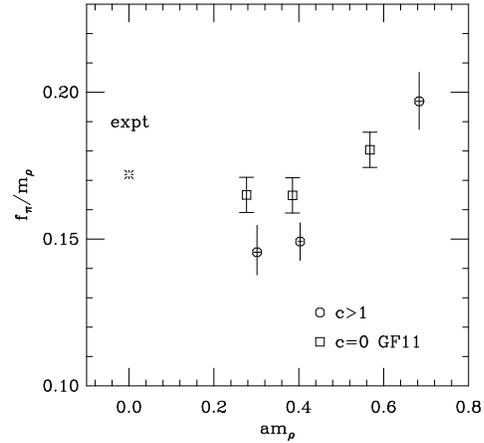}
\caption{$f_\pi$ in units of $m_\rho$.}
\label{fig:f_pi}
\end{figure}

Comparison of the pion decay constant values obtained with different
values of $c$ is complicated by systematic effects inherent in the
current normalisation prescription.  Given this caveat, we show our
tadpole-improved results along with those of GF11~\cite{gf11} in
Figure~\ref{fig:f_pi}. 
Both sets of data suggest scaling may set in above $\beta=6.0$.

\begin{figure}[htb]
\epsfxsize=0.4\textwidth
\epsffile{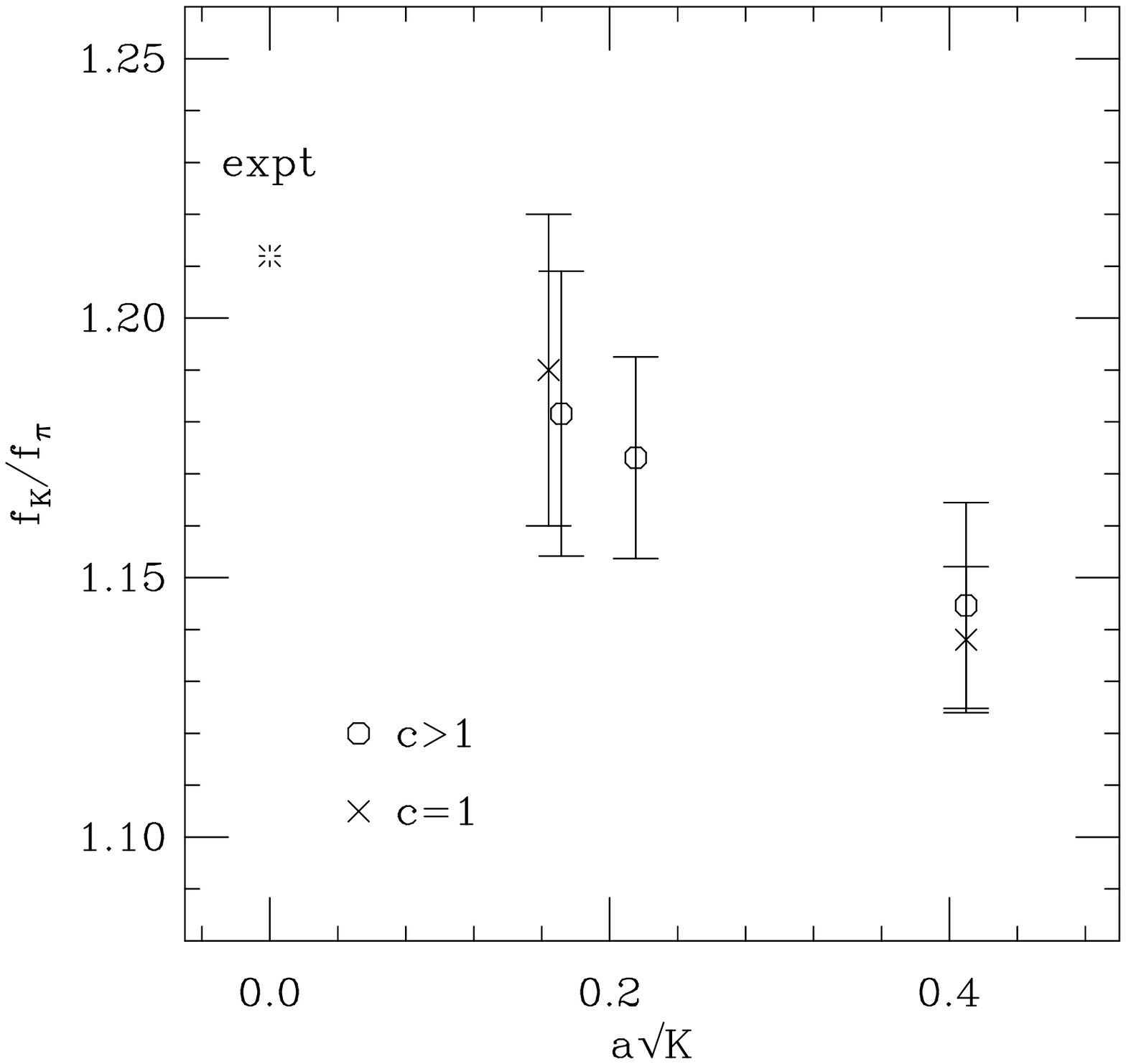}
\caption{$f_K/f_\pi$ versus lattice spacing.}
\label{fig:f_K_to_f_pi}
\end{figure}

The ratio of decay constants, $f_K/f_\pi$, is independent of the
current normalisation and so may be determined more reliably. Our
results, shown in Figure~\ref{fig:f_K_to_f_pi}, are insensitive to
whether $c=1$ or $1/u_0^3$. 
For $\beta\ge 6.0$ the ratio agrees with experiment, but a weak lattice
spacing dependence remains below this $\beta$ value.

\section{CONCLUSIONS}

Using a tadpole-improved SW fermion action in quenched QCD, we conclude
the following. 

1. For $\beta\ge 5.7$, $m_\rho/\sqrt{K}$ has a weaker dependence on the
lattice spacing (Figure~\ref{fig:m_rho}), $m_N/m_V$ and
$m_{K^\ast}/m_\rho$ scale (Figures~\ref{fig:ed_scaling} and
\ref{fig:kaons}), although the latter clearly shows that the strange
quark mass cannot be determined consistently.

2. Some quantities, such as $m_{K^\ast}/m_\rho$, $m_V^2-m_{PS}^2$, $J$
and $f_K/f_\pi$, are insensitive to whether $c$ has its tree-level or
tadpole-improved value (Figures~\ref{fig:kaons} and \ref{fig:f_K_to_f_pi}).

3. Provided tadpole-improved perturbation theory for $Z_A$ is reliable,
there is an indication that $f_\pi/m_\rho$ may scale for $\beta\ge 6.0$
(Figure~\ref{fig:f_pi}).

\end{document}